\documentclass[aps,preprint,showpacs]{revtex4-1}
\usepackage{amsfonts}
\usepackage{amsmath}
\usepackage{amssymb}
\usepackage{graphicx}
\usepackage{rotating}
\usepackage{hyperref}

\begin{document}

\title{Phase structure of cold magnetized quark matter within the SU(3)
NJL model}

\author{Ana G. Grunfeld$^{a,b}$, D\'ebora P. Menezes$^{c,d}$, Marcus B. Pinto$^c$ and Norberto N. Scoccola$^{a,b,e}$ }
\affiliation{$^a$CONICET, Rivadavia 1917, 1033 Buenos Aires, Argentina\\
$^b$ Department of Theoretical Physics, Comisi\'on Nacional de Energ\'ia
At\'omica, Av.Libertador 8250, 1429 Buenos Aires, Argentina\\
$^c$ Departamento de F\'{\i}sica CFM,
Universidade Federal de Santa Catarina, Florian\'{o}polis, SC CP. 476, CEP 88.040-900, Brazil\\
$^d$ Departamento de F\'isica Aplicada, Universidad de Alicante,
Ap. Correus 99, E-03080, Alicante, Spain\\
$^e$ Universidad Favaloro, Sol\ 453, 1078 Buenos Aires, Argentina}

\pacs{24.10.Jv, 25.75.Nq}

\begin{abstract}

The possible different phases of cold quark matter in the presence
of a finite magnetic field and chemical potential are obtained
within the SU(3) NJL model for two parameter sets often used in
the literature. Although the general pattern is the same in both
cases, the number of intermediate phases is parameter dependent.
The chiral susceptibilities, as usually defined, are different not only
for the $s$-quark as compared with the two light quarks, but also
for the $u$ and $d$-quarks, yielding non identical crossover lines
for the light quark sector.
\end{abstract}

\maketitle

\section{Introduction}

The study of the QCD phase diagram, when matter is subject to
intense external magnetic fields has been a topic of intense
investigation recently. The fact that magnetic fields can reach
intensities of the order of $B \sim 10^{19}$ G or higher in
heavy-ion collisions \cite{kharzeev} and up to $10^{18}$ G in the
center of magnetars \cite{magnetars,njlB} made theoretical
physicists consider matter subject to magnetic field both at high
temperatures and low densities and low temperatures and high
densities. Most effective models foresee that at zero chemical
potential a crossover transition is obtained at pseudo critical
temperatures that increase with an increasing  magnetic field, a
behavior which is contrary to the one found in lattice QCD (LQCD)
calculations \cite{baliJHEP2012,ruggieri2,Fraga:2012rr}.
Possible explanations for this discrepancy have been recently given in
Refs. \cite {prl,kojo,endrodi}.
As  LQCD
calculations are not yet in position to describe the whole $T-\mu$
plane, further investigations with effective models have been
developed towards a better understanding of the behavior of the
quark condensates \cite{marcio2013}, in the search for coexistent
chemical potentials  at sub-critical temperatures \cite{prdrapid}
as well as the existence and possible location of the critical end
point \cite{prdrapid,pedro2013}. In the case of magnetized quark
matter some interesting  results were obtained from these
investigations. Namely,  the first order segment  of the
transition line becomes longer as the field strength increases so
that a larger coexistence region for hadronic and quark matter
should be expected for strong magnetic fields affecting the
position of the (second order) critical point where the first
order transition line terminates.

In Ref. [11] it was observed that at  sub-critical temperatures the coexistence chemical potential ($\mu_c$) initially decreases
with increasing values of the magnetic fields but the situation gets reversed when extremely high fields (higher than the cut off scale) are considered so that $\mu_c$ oscillates
around the B = 0 value for magnetic fields within the $10^{17} - 10^{20}$ G range. Together, these effects have interesting consequences
for quantities which depend on the details of the coexistence region such as the surface tension as recently discussed in
Ref.\cite {andre}. Note that other physical
possibilities such as the isospin and strangeness content of the
system, the presence of a vector interaction \cite {robson}, and
the adopted model approximation within a particular
parametrization  may influence these results mostly in a
quantitative way.

As it is well known, in the absence of a
magnetic field dynamical chiral symmetry breaking (DCSB) occurs,
within four fermion theories, when the coupling ($G$) exceeds a
critical value ($G_c$), at least in the $2+1$ d and $3+1$ d cases.
However, when $B\ne 0$ DCSB  may occur even when the coupling is
smaller than $G_c$.  This effect, which is related to a
dimensional reduction induced by $B$, is known as magnetic
catalysis (MC). It was first observed in the $2+1$ d Gross-Neveu
model \cite {klimenkoGN} and then explained in Ref. \cite
{miransky} (see Ref. \cite {reviews} for recent reviews).
Following its discovery, Ebert, Klimenko, Vdovichenko and
Vshivtsev \cite {ebert} recognized that MC associated to the
filling of Landau levels could lead to more exotic phase
transition patterns as a consequence of the induced magnetic
oscillations. To confirm this assumption these authors have
considered a wide range of coupling values for  the two flavor
Nambu-Jona-Lasinio model (NJL) in the chiral case \cite{njl}. As
expected, they have  observed many phase structures as a function
of the  chemical potential: an infinite number of massless
chirally symmetric phases, a cascade of massive phases with broken
chiral invariance and tricritical points were also obtained.
Recently, these seminal works have been extended by a more
systematic, and numerically accurate analysis of the two flavor
case considering different model parametrizations identified by
the vacuum value of the dressed quark mass in the absence of an
external magnetic field $M_0$ \cite{pablo}. In that reference
other relevant physical quantities, such as susceptibilities, have
been considered in order to produce a phase diagram for cold
magnetized quark matter. Although the more complex transition
patterns show up for rather low values of the dressed quark mass $M_0
\simeq 200 \, {\rm MeV}$,
even with more canonical values of the model parameters leading to $M_0 \simeq 300 - 400 \, {\rm MeV}$, %one may observe
more than one first order phase transition, which is signaled when the thermodynamical potential develops two degenerate  minima at different values
of the coexistence chemical potential, is found.
 We point out that this fact has also been recently observed to arise within another
effective four fermion theory described by the $2+1$ d Gross-Neveu model \cite{enois}.
In general, {\it weak} first order transitions can be easily missed in a numerical evaluation due to the fact that
the two degenerate minima appear almost at the same location being separated by a tiny potential barrier so that
their study requires extra care. Physically, this corresponds  to a situation where  two different (but almost identical)
densities coexist at the same chemical potential, temperature and pressure.  Also, since these shallow minima are separated
by a low potential barrier, one may also expect  the surface tension to be small
in this case \cite {randrup}.

At this point it is important to recall that strangeness is
generally believed to be of great relevance for the physics of
quark stars and the heavy ion collisions and hence cannot be
disregarded. For instance, in astrophysical applications the
magnetic oscillations studied in Refs. \cite{ebert,pablo} may
influence the equation of state (EoS) which is the starting point
as far  as the prediction of observables, such as the mass and
radius of a compact star, is concerned. Therefore, and as a step towards the full understanding of the role played by strangeness in these
physical situations, in the present work we extend the detailed analysis of cold quark matter recently
performed with the two-flavor version to the three-flavor version of the NJL model which is described in terms of two canonical sets of input
parameters. In the next section we obtain the pressure for the
three flavor NJL and, in Sec. III, we present our numerical
results. Our final remarks are presented in Sec. IV.

 \section{Formalism}

We consider the SU(3) NJL Lagrangian density which includes a
scalar-pseudoscalar interaction and the t\'{}Hooft six-fermion
interaction\cite{hat} and is written as:
\begin{equation}\label{lf}
    \mathcal{L}=\bar{\psi}\left( i\ \rlap/\!D
    -\hat{m}_{c} \right)\psi+ G\sum_{a=0}^{8}\left[ \left( \bar{\psi}\lambda_{a}\psi \right)^{2}
    +\left( \bar{\psi}i\gamma_{5}\lambda_{a}\psi \right)^{2} \right]
-K \left( d_{+}+d_{-}\right)\,,
\end{equation}

\noindent where $G$ and $K$ are coupling constants, $\psi
=\left(u,d,s\right)^{T}$ represents a quark field with three
flavors, $d_{\pm}=det\left[ \bar{\psi}\left(
1\pm\gamma_{5} \right)\psi \right]$,
$\hat{m}_{c}=\mathrm{diag}\left( m_{u},m_{d},m_{s} \right)$ is the
corresponding (current) mass matrix, $\lambda_{0}=\sqrt{2/3}\,I$,
where $I$ is the unit matrix in the three flavor space, and
$0~<~\lambda_{a}~\leq~8$ denote the Gell-Mann matrices. The
coupling of the quarks to the electromagnetic field ${\cal A}_\mu$
is implemented through the covariant derivative
$D_{\mu}=\partial_\mu - i \hat q {\cal A}_{\mu}$ where $\hat q$
represents the quark electric charge matrix $\hat q
=\mathrm{diag}\left( q_{u},q_{d},q_{s} \right)$ where $q_u/2 = -
q_d = - q_s = e/3$. In the present work we consider a static and
constant magnetic field in the z direction, ${\cal
A}_\mu=\delta_{\mu 2} x_1 B$. In the mean-field approximation the
Lagrangian density Eq.(\ref{lf}) can be written as \cite{njlB}
\begin{equation}\label{lf_mfa}
  \mathcal{L}^{\mathrm{MFA}} = \bar{\psi}\left( i\ \rlap/\!D-\hat{M}\right)\psi
      -2G\left( \phi_{u}^{2}+\phi_{d}^{2}+\phi_{s}^{2} \right)+4K\phi_{u}\phi_{d}\phi_{s}\,,
\end{equation}

\noindent
where $\hat{M}=\mathrm{diag}\left( M_{u},M_{d},M_{s} \right)$ is a matrix with elements defined by the
dressed quark masses which satisfy the set of three coupled gap equations
\begin{eqnarray}
f_u(M_u,M_d,M_s) &=&  M_{u}- m_u + 4 G\ \phi_{u}- 2K\ \phi_{d}\phi_{s} = 0 \ ,
\nonumber\\
f_d(M_u,M_d,M_s) &=&  M_{d}- m_d + 4 G\ \phi_{d}- 2K\ \phi_{s}\phi_{u} = 0 \ ,
\nonumber\\
f_s(M_u,M_d,M_s) &=&  M_{s}- m_s + 4 G\ \phi_{s}- 2K\ \phi_{u}\phi_{d} = 0 \ .
\label{gapeqs}
\end{eqnarray}
In Eqs.(\ref{lf_mfa},\ref{gapeqs}), $\phi_{f}$ is the quark condensate associated to each flavor which contains three different
terms: the vacuum, the magnetic and the in medium one. At vanishing temperatures these contributions read
\begin{equation}
 \phi_{f}=\langle{\bar \psi}_{f} \psi_{f}\rangle= \phi_{f}^{\mathrm{vac}}+\phi_{f}^{\mathrm{mag}}+\phi_{f}^{\mathrm{med}}
\end{equation}
where
\begin{eqnarray}
\phi_{f}^{\mathrm{vac}} &=& -\frac{N_c  M_{f}}{2\pi^{2}} \left[\Lambda \epsilon_\Lambda-{M_{f}^{2}}\ln \left ( \frac{\Lambda+ \epsilon_\Lambda}{{M_{f} }} \right ) \right ] \mbox{,}
\nonumber \\[2.mm]
\phi_{f}^{\mathrm{mag}} &=& -\frac{N_{c} M_{f}|q_{f}| B}{2\pi^{2}}\left [ \ln \Gamma(x_{f}) -\frac{1}{2} \ln (2\pi) \right .
+ \left . x_{f} -\frac{1}{2} \left ( 2 x_{f}-1 \right )\ln (x_{f}) \right ] \mbox{,}
\nonumber \\[2.mm]
\phi_{f}^{\mathrm{med}} &=& \frac{N_{c}}{2\pi^{2}} \ M_f \ |q_{f}|B \sum_{\nu=0}^{\nu^{max}_f} \alpha_\nu \
 \ln\left[ \frac{\mu+\sqrt{\mu^{2}-s_{f}(\nu,B)^{2}}}{s_{f}(\nu,B)} \right] \mbox{,}
\label{conds}
\end{eqnarray}
\noindent
where $s_{f}(\nu,B)=\sqrt{M_{f}^2+2|q_{f}|B\nu}$, $\epsilon_\Lambda=\sqrt{\Lambda^2 + M_{f}^{2}}$
with $\Lambda$ representing a non covariant
ultraviolet cutoff \cite{ebert} while
$x_f = M_{f}^{2}/(2 |q_{f}| B)$ and $\mu$ is the quark chemical potential.
Note that, for simplicity,  in the present work we consider the case of
symmetric matter where all three quarks carry the same chemical potential.
In $\phi^{med}_f$, the sum is over the Landau levels (LL´s), represented by $\nu$, while $\alpha_\nu=2-\delta_{\nu 0}$ is a degeneracy factor
and $\nu^{max}_f$ is the largest integer that satisfies $\nu^{max}_f \le (\mu^2 - M^2_f)/(2 |q_f| B)$.

Then, within the mean-field approximation, the grand-canonical thermodynamical potential cold and dense strange quark matter in the presence of an external magnetic field can be written as
\begin{equation}\label{p_mfa}
    \Omega=-\left(\theta_{u}+\theta_{d}+\theta_{s}\right)+2G\left( \phi_{u}^{2}+\phi_{d}^{2}+\phi_{s}^{2} \right)-4K\ \phi_{u}\phi_{d}\phi_{s}\,,
\end{equation}
where $\theta_{f}$ gives the contribution from the gas of quasi-particles and can be written as the sum
of 3 contributions,
\begin{eqnarray}
\theta^{\mathrm{vac}}_{f} &=&- \frac{N_c}{8\pi^2} \ \left \{ M_f^4 \ln \left[\frac{(\Lambda+ \epsilon_\Lambda)}{M_f} \right]
-\epsilon_\Lambda \, \Lambda\left(\Lambda^2 +  \epsilon_\Lambda^2 \right ) \right \} \mbox{,}
\nonumber \\[2.mm]
\theta^{\mathrm{mag}}_{f} &=&
\frac {N_{c} }{2 \pi^2} \ (|q_{f}|B)^2 \ \left [ \zeta^{(1,0)}(-1,x_{f}) -  \frac {1}{2}( x_{f}^{2} - x_{f}) \ln x_{f} +\frac {x_{f}^{2}}{4} \right ] \mbox{,}
\nonumber \\[2.mm]
\theta^{\mathrm{med}}_{f}&=&
\frac{N_{c}}{4\pi^{2}} \ |q_{f}|B \ \sum_{\nu=0}^{\nu^{max}_f} \alpha_\nu \left[ \mu\sqrt{\mu^{2}-s_{f}(\nu,B)^{2}}
\right.\nonumber\\
& &
\left. \qquad \qquad \qquad
-s_{f}(\nu,B)^{2}\ln\left( \frac{\mu+\sqrt{\mu^{2}-s_{f}(\nu,B)^{2}}}{s_{f}(\nu,B)} \right) \right]  \mbox{,}
\label{PmuB}
\end{eqnarray}

\noindent
where $\zeta^{(1,0)}(-1,x_f)= d\zeta(z,x_f)/dz|_{z=-1}$ with $\zeta(z,x_f)$ being the Riemann-Hurwitz zeta function.

\begin{table}[h]
\centering
\begin{tabular}{cccccc}
  \hline  \hline
\hspace*{.5cm} Parameter set \hspace*{.5cm}   & \hspace*{.5cm} $\Lambda$ \hspace*{.5cm} &
\hspace*{.5cm} $G\Lambda^2$ \hspace*{.5cm}    & \hspace*{.5cm} $K\Lambda^5$ \hspace*{.5cm} &
\hspace*{.5cm} $m_{u,d}$ \hspace*{.5cm} & \hspace*{.5cm} $m_{s}$ \hspace*{.5cm} \\
         &    MeV    &          &   &  MeV   &  MeV  \\ \hline
Set 1 \cite{hat2}& 631.4 & 1.835 & 9.29 & 5.5 & 135.7 \\
Set 2 \cite{reh} & 602.3 & 1.835 & 12.36 & 5.5 & 140.7 \\
\hline
  \hline
\end{tabular}
\caption{\label{pnjl} Parameter sets for the NJL SU(3) model.}
\end{table}

We next use the formalism given above to identify the different phase structures
of magnetized quark matter. To obtain the critical chemical potential at given
$eB$ we proceed as it follows. In the case of first
order phase transitions, we have used the same prescription as in \cite{prdrapid},
i.e., we have calculated the thermodynamical potential as a function of the dressed
quark masses and then searched for two degenerate minima. In the case of crossover
transitions its position is identified by the peak of the chiral susceptibility.
However, differently from the standard SU(2) NJL model with maximum flavor mixing,
within the present version of the SU(3) NJL model the $u$ and $d$ dressed
quark masses, as well as the corresponding condensates, are not necessarily the same.
For this reason, we have defined different susceptibilities for each flavor
$\chi_f= d \phi_f / d m_f$.
The peak of these susceptibilities are used in the next section to identify
possible crossover transitions.

\section{Results and discussion}

In this section we present and analyze the results of our numerical investigations.
These were performed by solving the set of three coupled gap equations given
in Eq.(\ref{gapeqs}) for different values of the chemical potential and
magnetic fields. In order to analyze the dependence of the results on
the model parameters we consider two widely used SU(3) NJL model
parametrizations. Set~1 corresponds to that used in Ref.\cite{hat2} while
Set 2 to that of Ref.\cite{reh}. The corresponding model parameters
are listed in Table \ref{pnjl}. It should be stressed that, contrary
to what happens in the case of the standard SU(2) NJL model with maximum
flavor mixing \cite{njl,pablo}, for these parametrizations the difference
between the $u$ and $d$ quark electric charges induces a splitting between the
$u$ and $d$ dressed quark masses. Nonetheless, we have found that, in general,
this splitting is quite small. Of course, due to the larger value of the
associated current quark mass, $M_s$ is always larger than $M_u$ and $M_d$.

We stress that, for a given value of $\mu$ and $B$,
the set of
coupled gap equations might have several solutions. Obviously, the physical
solution
is the one that leads to the lowest value of the thermodynamical potential
Eq.(\ref{PmuB}). Therefore, it is important to make sure that one does
not miss any relevant solution when solving Eqs.(\ref{gapeqs}). In order
to do that we have proceeded as follows: given a value of $M_u$
the set of equations $\{f_d,f_s\}$ were used to numerically
determine the corresponding values of $M_d$ and $M_s$. These values
were then inserted in the remaining gap equation which could be now
considered as a function of the single variable $M_u$, i.e.
$f_u\left(M_u,M_d(M_u),M_s(M_u)\right)$. Varying $M_u$ within a conveniently
selected range of values one could at this point determine all the solutions of the coupled
system of gap equations by finding all the values of $M_u$ at which this function
vanishes. Of course, one has to be careful with a possible caveat that this method
can have: it could happen that for a given value of $M_u$ the set of equations
$\{f_d,f_s\}$ might have more that one solution. We have verified that, for
the model parametrizations used in this work, which imply a rather strong flavor mixing
leading to $M_u \approx M_d$, this situation did not arise for any of the
values of $\mu$ and $B$ considered.

\begin{figure}[ht]
\begin{center}
\includegraphics[angle=0,width=0.90\linewidth]{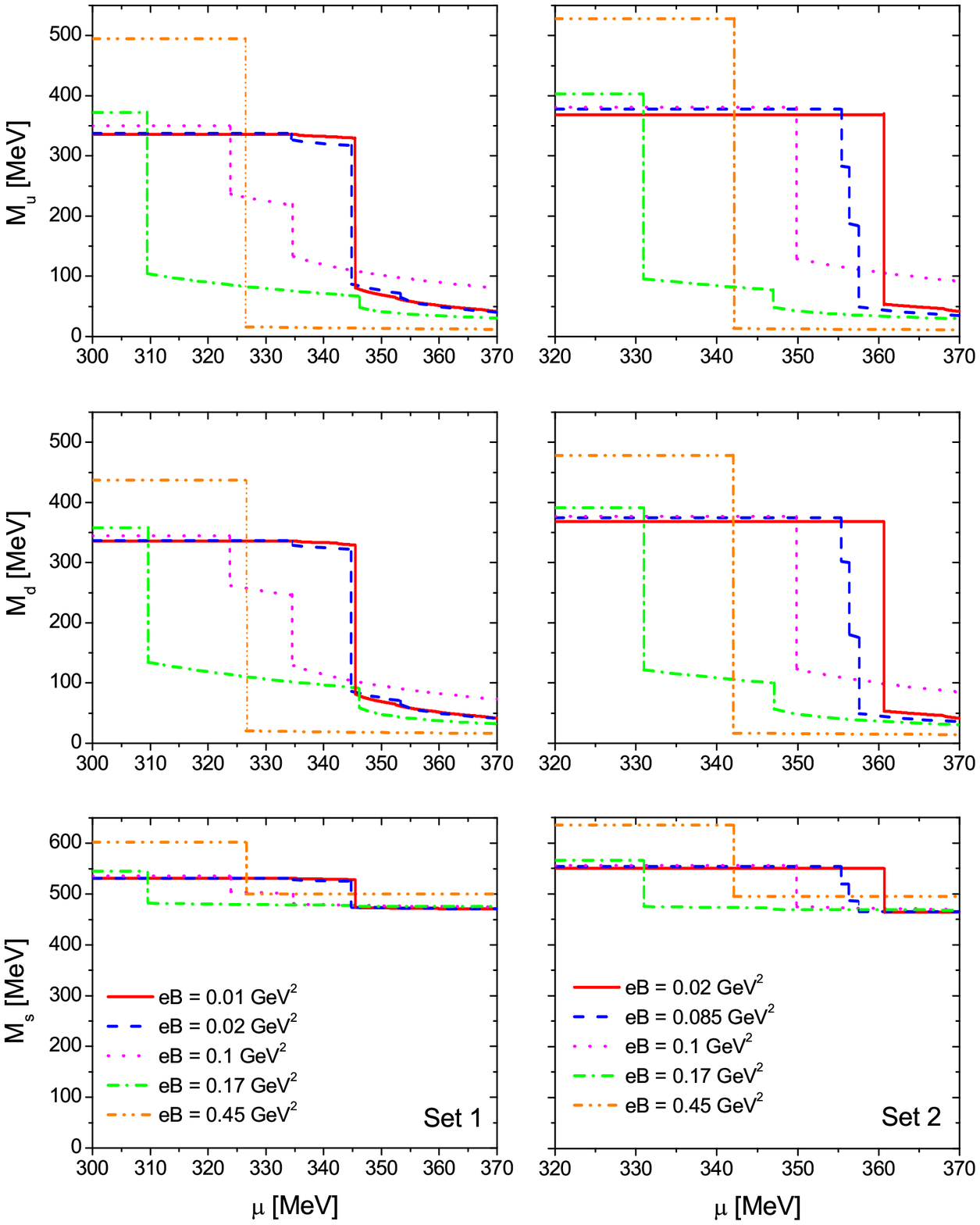}
\end{center}
\caption{Dressed quark masses as functions of chemical potential for different values of the
magnetic field.}
\label{Mvsmu}
\end{figure}

We start by analyzing the behavior of the quark constituent masses
as functions of the chemical potential for several representative values
of the magnetic field shown in Fig.\ref{Mvsmu}. Note that here, and in
what follows, we  use natural units recalling the reader that $eB = 1\ \mbox{GeV}^2$
corresponds to $B= 1.69 \times 10^{20}$ G. We consider first the
situation for Set~1 (left panels) starting by the lowest chosen value of
magnetic field $eB= 0.01 \ \mbox{GeV}^2$ (full red line). As we increase $\mu$ we
see that up to certain value $\mu_{c1}=335.3 \ \mbox{MeV}$ the dressed masses stay constant.
At that point we can observe a (tiny) sudden drop in the masses
corresponding to a first order phase transition
which goes from the fully chirally broken phase, where the masses are $\mu-$independent,
to a less massive one where the masses are $\mu-$dependent.
As we continue to increase the chemical potential there is a second tiny
jump in the masses (somewhat more easily observed in the plot for $M_d$)
at $\mu_{c2}=342.3 \ \mbox{MeV}$. Increasing $\mu$ further we reach $\mu_{c3}=345.4 \ \mbox{MeV}$
where a new, in this case much larger, drop in the masses occur. After this
point the dressed $u-$ and $d-$quark masses are much smaller that their vacuum
values indicating that light quark sector is in the fully chirally restored phase, namely
that if we were to set $m_u=m_d=0$ (i.e. chiral case) the associated dressed
masses would vanish. To identify the different phases it is convenient to adopt
the notation of Refs.\cite{njl,pablo}. Thus, the fully chirally
broken phase in which the system is for $\mu < \mu_{c1}$ is denoted by B.
The massive phases in which $M_f$ depends on $\mu$ are denoted $\mbox{C}_i$ phases.
%Thus,
Hence, the system is in the $\mbox{C}_0$ phase if %for
$\mu_{c1} < \mu_{c2}$ and in the
$\mbox{C}_1$ phase if $\mu_{c2} < \mu_{c3}$. The difference between these
two phases
is that in the $\mbox{C}_0$ only the $u-$ and $d-$quark lowest Landau levels (LLL's) are populated
while in $\mbox{C}_1$ the $d-$ quark first LL (1LL) also is. It is important to remark that within the range of values of the chemical potential
considered in this work no $s-$quark LL is ever populated since this would
require larger values of $\mu$. Moreover, the reason why the
$d$-quark 1LL is populated without a simultaneous population of the
$u$-quark 1LL is due to fact that $|q_d| = 2 |q_u|$. For
$\mu > \mu_{c3}$ the system is in one of the chirally restored phases $\mbox{A}_i$,
which differ between themselves in the number of light quark LL's
which are populated. The transitions between these phases
correspond to small jumps in the masses (i.e. first order transitions)
which in Fig.\ref{Mvsmu} are hardly seen in this case.
Considering next the case $eB= 0.02 \ \mbox{GeV}^2$ (blue dashed line)
we see that at basically the same value of $\mu_{c1}$ as the one given above
there is a first order transition from the B phase to the $\mbox{C}_0$ phase.
However, in this case no sign of a transition to the $\mbox{C}_1$ phase is found as
$\mu$ increases.
In fact, the following transition corresponds to a big jump in the light quark dressed
masses which is associated to the transition between the $\mbox{C}_0$ and one of the
chiral symmetry restored phases $\mbox{A}_i$. It should be noted that this
happens at a critical chemical potential which is slightly smaller than the
value of $\mu_{c3}$ quoted above. As $\mu$ is further increased  consecutive
first order transitions between $\mbox{A}_i$ occur. Note that the first
of them can quite clearly observed in this case.
Turning to the case $eB= 0.10 \ \mbox{GeV}^2$ (magenta dotted lines)
we see that the overall behavior is similar to that of $eB= 0.01 \ \mbox{GeV}^2$,
except for the fact that in the present case the size of the first and second jumps
in the masses are quite similar and that they occur at lower values of $\mu$.
The situation for $eB= 0.17 \ \mbox{GeV}^2$ (green dash dotted lines) is
somewhat peculiar. After a first large jump in the masses (occurring at an
even lower
value of $\mu$ than the previous cases) they decrease continuously for a rather
wide interval of values of $\mu$ which ends with a transition which
is characteristic of those between $\mbox{A}_i$ phases. Whether in that
intermediate interval the system is always in the same phase (of $\mbox{A}_i$
type)
or it stays first in the $\mbox{C}_0$ phase performing at some point a crossover
transition to a $\mbox{A}_i$ type phase is a question that requires further
analysis and %will be
is addressed below. Finally, for $eB= 0.45 \ \mbox{GeV}^2$ (orange dash dot dotted lines) the behavior
of the system as $\mu$ increases becomes much simpler. There is one single first
order phase transition which connects the B and $\mbox{A}_0$ phases. Note,
however,
that this transition occurs at a higher chemical potential as compared to that
required in the previous case to induce a transition from the B phase.
Turning our attention to the results concerning Set 2 (right panels in
Fig.\ref{Mvsmu}) we observe that, although for $eB= 0.45 \ \mbox{GeV}^2$
(orange dash dot dotted lines) the behavior is very similar as the
corresponding one for Set~1,
at low values of $eB$ there are significant differences. For example
for $eB= 0.02 \ \mbox{GeV}^2$ (full red lines) the first transition already
connects the B phase with one of the $\mbox{A}_i$ ones, i.e. there is no sign
of an intermediate $\mbox{C}_0$ here. As it turns out such a phase
only exists for a narrow interval of values of $eB$ of which
we take $eB= 0.085 \ \mbox{GeV}^2$ (dashed blue line) as a typical example.

\begin{figure}[ht]
\begin{center}
\includegraphics[angle=0,width=0.70\linewidth]{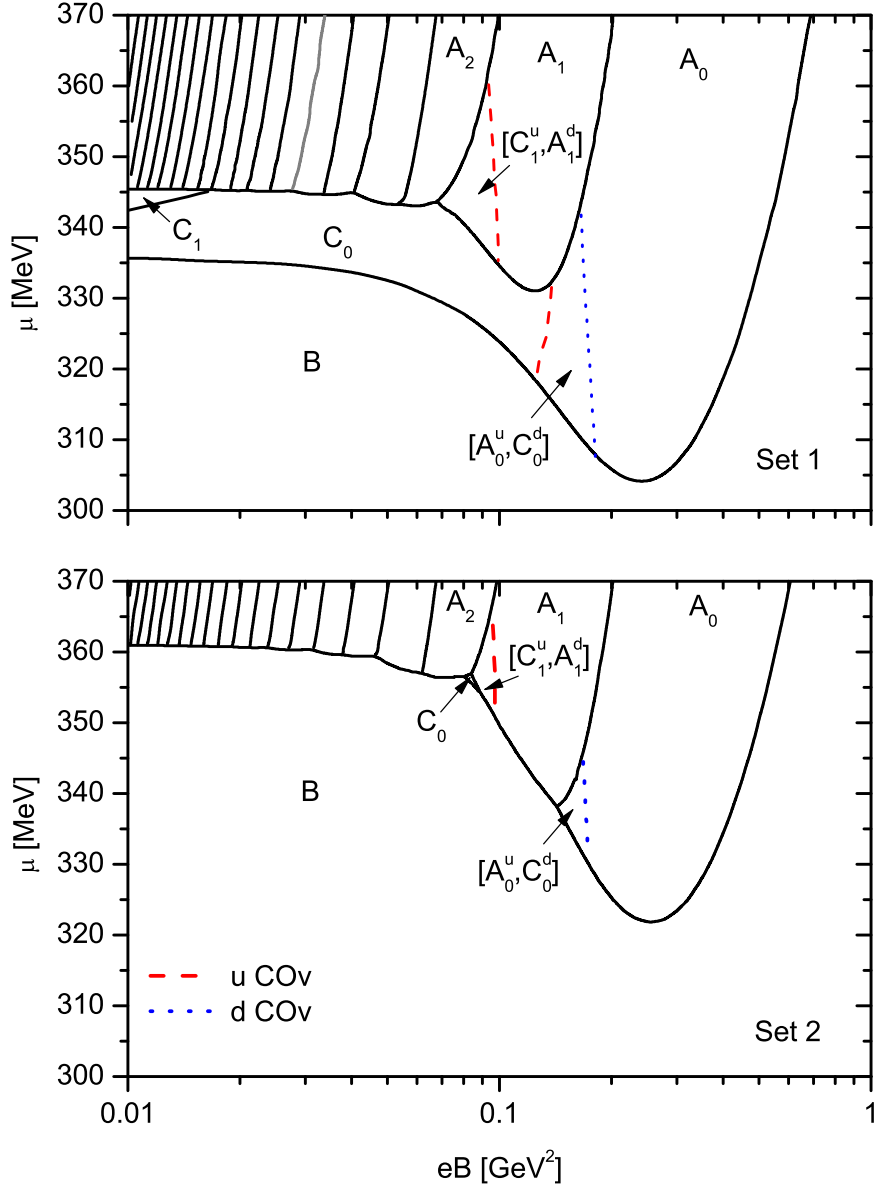}
\end{center}
\caption{Phase diagrams in the $eB-\mu$ for the set of parameters defined in Table I.}
\label{fig_phase}
\end{figure}

In Fig.\ref{fig_phase} we show the $eB-\mu$ phase diagrams obtained with both
parameter sets. The full lines correspond to first order phase transitions
while the dashed and
dotted ones to smooth crossovers. As mentioned above the later typically connect
some of the $C_i$ and $A_i$ phases and their determination requires a detailed
analysis. In the first place we should stress that, as well known, there is
not a unique way
to define a crossover transition. In the case of SU(2) cold quark matter under
strong magnetic fields this issue was discussed in some detail in Ref.\cite{pablo}.
Following that reference we base our analysis on the chiral susceptibilities
introduced at the end of the previous section. In particular, we define the
crossover
transition line as the ridge occurring in the chiral susceptibility when regarded as a
two dimensional function of $eB$ and $\mu$. Mathematically, it can be defined by using for
each value of the susceptibility (starting from its maximum value in the given region)
the location of the points at which the gradient in the $eB-\mu$ plane is smaller.
As remarked in Ref.\cite{pablo}, this definition must be complemented with the condition
that on each side of the curve there should exist at least one region such that there is
a maximum in the susceptibility for an arbitrary path connecting both regions.
It is important to note  that, differently from the $SU(2)$ case discussed
in that reference where one single chiral susceptibility can be defined for the two light flavors,
the values of  $\chi_u$ and $\chi_d$ at a given point in the $eB-\mu$ plane are in general different
in the present case \cite{fot}.
Therefore, there is no reason why there should be identical crossover lines for the two
light quark sectors. In fact, and in contrast to what happens with the first order
transitions which always coincide, the result of our analysis indicate that for the
parametrizations considered this is never the case. As a consequence of this
there might be regions in the $eB-\mu$ plane where the $u-$quark sector is in
a $\mbox{C}_i$ phase and the $d-$quark sector in a $\mbox{A}_i$ one and viceversa.
In Fig.\ref{fig_phase}, those regions have been indicated by including the
notation for each of the
corresponding phases (i.e. one for each light quark sector) between brackets.
Thus, for example, $[\mbox{A}^u_0,\mbox{C}^d_0]$ corresponds to a region in which the
$u-$ quark sector is in the $\mbox{A}_0$ phase and the $d-$ quark sector $\mbox{C}_0$.
From Fig.\ref{fig_phase} it is clear that the most remarkable difference
between the phase diagrams associated to the two parametrizations considered
concerns the regions in which the phases $\mbox{C}_i$ exist. For Set~1 the $\mbox{C}_0$
phase covers a rather large area of the plane, which in the $eB$ direction
extends from very low values up to $eB \simeq 0.15 \mbox{GeV}^2$ where
it has a smooth crossover boundary with the phase $A_0$. Note that such
a boundary is somewhat different for the two light quark sectors giving
rise to an intermediate $[\mbox{A}^u_0,\mbox{C}^d_0]$ region. Moreover,
for this parameter set a small region of $\mbox{C}_1$ phase exists for low values
of $eB$. In the case of Set 2, however, the phase $\mbox{C}_0$ only exists in
a small triangular region surrounded by first order transition lines although
a small $[\mbox{A}^u_0,\mbox{C}^d_0]$ region is also present.
Another point that it is interesting to address regards the similarities
between the present phase diagrams and those reported in Ref.\cite{pablo}
for the SU(2) case with maximum flavor mixing. In fact, the phase diagram
obtained for Set~1 has strong similarities to that shown for $M_0=340 \ \mbox{MeV}$
shown in Fig.12 of that reference. Moreover, that of Set 2 appears
to correspond to one somewhere in between those of $M_0=360 \ \mbox{MeV}$
and $M_0=380 \ \mbox{MeV}$ of that figure. Interestingly, the vacuum values
of the $u-$ and $d-$ dressed quark masses in the vacuum and in the absence of
a magnetic field are $M_u=M_d=336(368)$~MeV for Set~1 (Set~2). Thus, it
appears that even in the $SU(3)$ case under consideration the general
features of the $eB-\mu$ diagram are dictated to a great extent by
the values of light quark dressed quark masses in the vacuum and in
the absence of a magnetic field.

\begin{figure}[ht]
\begin{center}
\begin{tabular}{ll}
\includegraphics[angle=0,width=0.90\linewidth]{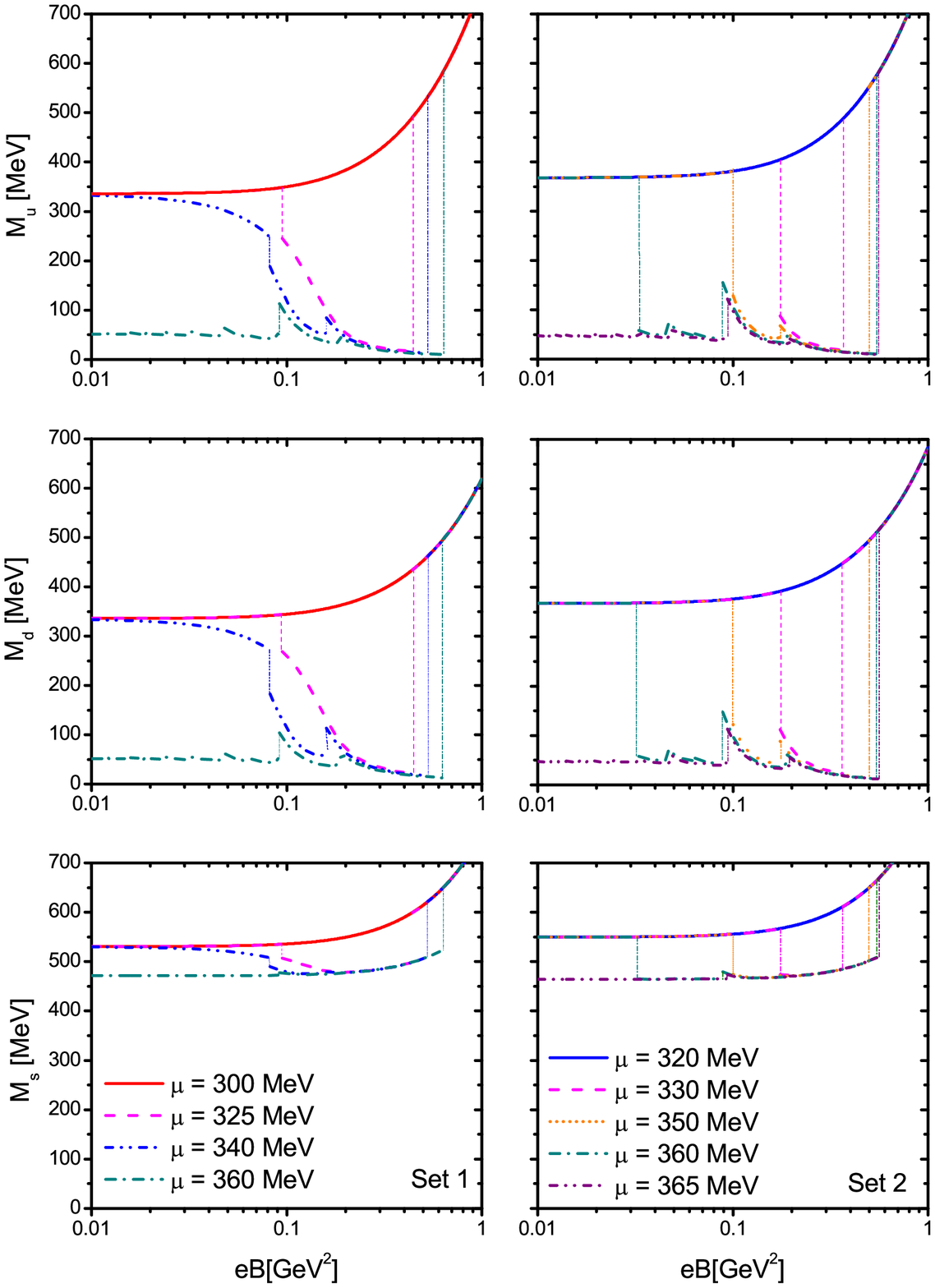} \\
\end{tabular}
\end{center}
\caption{Dressed quark masses as functions of magnetic field for different values of the
chemical potential.}
\label{MvseB}
\end{figure}

We end this section by analyzing in the context of the present SU(3) NJL model
the magnetic catalysis (MC) effect mentioned in the Introduction and how
this effect is modified by the presence of finite chemical potential leading,
for example, to the existence of the so-called inverse magnetic catalysis (IMC) as it has
been recently discussed
in the literature \cite{Preis:2010cq}. Actually, the later is usually related
to a decrease of the critical chemical potential at
intermediate values of the magnetic fields. Such a phenomenon is clearly
observed in the
phase diagrams displayed in Fig.\ref{fig_phase}. In fact, we see that after
staying fairly constant
up to $eB \simeq 0.05 \ \mbox{GeV}^2$ the lowest first order transition line bends down reaching a
minimum at $eB \simeq 0.2 - 0.3 \ \mbox{GeV}^2$ after which it rises
indefinitely with the magnetic field. This implies that, in general, there is
some interval of values
of the chemical potential for which an increase of the magnetic field at
constant $\mu$ causes first a transition from the massive phase B to a less
massive phase ($\mbox{C}_i$ or $\mbox{A}_i$)
and afterwards from the massless phase $\mbox{A}_0$ back to massive phase B.
To address these issues we display in Fig.\ref{MvseB} the behavior of the
masses as function of magnetic
fields for several chemical potentials, and our two parameter sets.
In particular, the complex phase structure for Set~1 (left panels) accounts
for the different possible
behaviors depending on the chemical potential. For $\mu = 300$ MeV (red full
line), the system is in the
B phase for the whole range of magnetic fields, and the MC effect is clearly seen.
For $\mu = 325$ MeV (magenta dashed line), a similar behavior is seen, except for a middle section where
the system passes through a $\mbox{C}_0$ phase and an $\mbox{A}_0$ phase before returning to the vacuum phase again.
Note that when masses are plotted as functions of $eB$, the existence of a crossover transition from
$\mbox{C}_0$ to $\mbox{A}_0$ becomes more noticeable. As already discussed, a detailed analysis
shows that the associated critical magnetic field for $u-$quark sector is somewhat lower than
that for $d-$quark sector. In this
region of the curve, as well as in the rest of the following curves, the effect of IMC
is also present. In fact, as already remarked in Ref.\cite{pablo}, we can conclude that
within the $C_i$ and $A_i$ phases the dressed light quark masses are
basically decreasing functions of the magnetic field, while MC
occurs principally in the B phase. In particular, for $\mu = 340$ MeV (blue dot dotted lines),
the phase remains in $\mbox{C}_0$ for a significant range of magnetic fields and the mass decreases continuously.
Finally, for $\mu = 360$ MeV (green dash dotted lines) at low and medium magnetic fields the system goes
from a $\mbox{A}_i$ phase to another one with $i'=i-1$ as the magnetic fields increases, the transition between them
being signalled by the peaks in the dressed masses. Eventually, for
sufficiently large magnetic fields,
it has a first order transition to the B phase. As shown in the right panels of Fig.\ref{MvseB}, for
Set 2 the situation is somewhat simpler. This is, of course, related to the absence of extended $\mbox{C}_i$ regions
in the associated phase diagram.

\section{Final remarks}

In the present work we have revisited the phase structure of the magnetized cold quark
matter within the framework of the SU(3) NJL model for two parameter sets often used in
the literature\cite{Buballa}. Although the general pattern is similar, the quantitative
results are certainly parameter dependent, as in the case of the SU(2) NJL \cite{ebert,pablo}.
We have checked that Set~1, i.e. the one leading to lower vacuum values for
the dressed quark masses in the absence of a magnetic field, presents a richer phase
diagram, with more intermediate phases than Set 2. This is a reflex of the number of
small jumps appearing in the quark dressed masses, which are related to
the number of filled Landau levels.
It is worth emphasizing that,
differently from the case of standard SU(2) NJL model with maximum flavor mixing studied
in Ref.\cite{pablo}, within the present version of the SU(3) NJL model the $u-$ and $d-$dressed
quark masses, as well as the corresponding condensates, are not necessarily
equal for the same chemical potential $\mu$ and magnetic field $B$.
As a consequence, three different susceptibilities (one for each flavor) can be defined
which, in principle, might bear peaks at different points. This points towards the possibility
of having a different crossover transition line for each of three quark flavors.
In fact, and in contrast to what happens with the first order
transitions which are always found to coincide, the result of our analysis indicate that for the
parametrizations considered this is always the case. Hence, the corresponding
phase diagrams turn out to have some (small) regions where the quarks of different flavor
are in different phases.

The phenomenon of inverse magnetic catalysis as defined in Ref.\cite{Preis:2010cq},
i.e. the decrease of the critical chemical potential at specific values of the magnetic
field intensity, is clearly observed within the present choice of parameters for the SU(3) NJL model.
In connection with this, we also note that the response of light quark dressed masses
to an increase of the magnetic at fixed $\mu$ depends on the region
of $eB-\mu$ phase diagram considered. On the one hand, the increase in light quark dressed masses
with magnetic field,
known as magnetic catalysis, is principally seen in the vacuum phase B, where chiral symmetry
is fully broken. On the other hand, phases where some light quark levels are populated ($\mbox{C}_i$ and $\mbox{A}_i$)
show a dominant decrease in the corresponding masses as magnetic field increased.

We conclude by noting that while in the present work we have restricted ourselves
to symmetric quark matter, the role played by charge neutrality and $\beta$-equilibrium
in the behavior of quark matter subject to strong magnetic fields is clearly a topic of
great interest \cite{njlB} in the study of magnetars.
As the existence of the critical end point is related to the amount of
different quark flavors in the system \cite{pedro2013}, the resulting
phase diagram will certainly be different, and at least at low temperatures,
it should be investigated.

\begin{acknowledgments}
This work was partially supported by CAPES, CNPq and FAPESC (Brazil),
by CONICET (Argentina) under grant PIP 00682 and
by ANPCyT (Argentina) under grant PICT-2011-0113.
\end{acknowledgments}

\begin{thebibliography}{999}
\bibitem{fot}
In fact, so it is $\chi_s$ but since such a quantity never presents a peak
for the values of $eB$ and $\mu$ considered in this work it will be ignored in what follows.
\end{thebibliography}
\end{document}